\documentclass[twocolumn,prl,showpacs,amsmath,amssymb]{revtex4}

\usepackage{graphicx}% Include figure files
\usepackage{dcolumn}% Align table columns on decimal point
\usepackage{bm}% bold math

\begin{document}

\title{Quantum Melting of Spin Ice: Emergent Cooperative Quadrupole and Chirality}% Force line breaks with \\

\author{Shigeki Onoda}
%\email{s.onoda@riken.jp}
\author{Yoichi Tanaka}
\affiliation{%
Condensed Matter Theory Laboratory, RIKEN, 2-1, Hirosawa, Wako 351-0198, Saitama, JAPAN
}%

\date{\today}% It is always \today, today,
             %  but any date may be explicitly specified

\begin{abstract}
A quantum melting of the spin ice is proposed for pyrochlore-lattice magnets Pr$_2TM_2$O$_7$ ($TM=$Ir, Zr, and Sn). The quantum superexchange Hamiltonian having a nontrivial magnetic anisotropy is derived
in the basis of atomic non-Kramers magnetic doublets.
The ground states exhibit a cooperative ferroquadrupole and pseudospin chirality, forming a magnetic analog of smectic liquid crystals. 
Our theory accounts for dynamic spin-ice behaviors experimentally observed in Pr$_2TM_2$O$_7$.
\end{abstract}

\pacs{Valid PACS appear here}% PACS, the Physics and Astronomy
                             % Classification Scheme.
%\keywords{Suggested keywords}%Use showkeys class option if keyword
                              %display desired
\maketitle

\begin{figure}
\begin{center}
\includegraphics[width=\columnwidth]{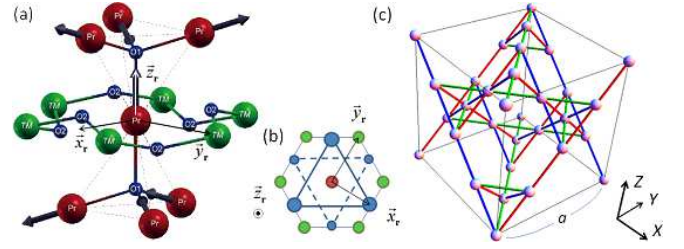}
\end{center}
\caption{(Color) (a) Pr$^{3+}$ ions (red) form tetrahedrons (dashed lines) centered at O$^{2-}$ ions (O1) (blue), and are surrounded by O$^{2-}$ ions (O2) (blue) in the $D_{3d}$ symmetry as well as by $TM$ ions (green). Each Pr magnetic moment (bold arrow) points to either of the two neighboring O1 sites. (b) The local coordinate frame $(\vec{x}_{\bm{r}},\vec{y}_{\bm{r}},\vec{z}_{\bm{r}})$ from the top. Upward and downward triangles of the O$^{2-}$ ions (O2) are located above and below the hexagon of the $TM$ ions. (c) The Pr pyrochlore lattice. The phase $\varphi_{\bm{r},\bm{r}'}$ in Eq.~(\ref{eq:H_eff}) depends on the color of the bonds. The global coordinate frame $(X,Y,Z)$ is also shown.}
\label{fig:crystal}
\end{figure}

It has been a great challenge to realize unconventional spin-liquid states in three-dimensional magnets. It is achieved by preventing a dipole long-range order (LRO) of magnetic moments, which requires appreciable quantum spin fluctuations and geometrical frustration of magnetic interaction~\cite{anderson:56,anderson,wen:89,palee:09}. The importance of the geometrical frustration is manifest as in pyrochlore systems~\cite{reimers:91,harris:97,ramirez:99,bramwell:01}. In particular, in the dipolar spin ice $R_2$Ti$_2$O$_7$ ($R$=Dy or Ho)~\cite{harris:97,ramirez:99,bramwell:01}, the rare-earth magnetic moment located at each vertex of tetrahedrons points either inwards (``in'') to or outwards (``out'') from the center (Fig.~\ref{fig:crystal} (a)). The nearest-neighbor ferromagnetic coupling mainly due to the magnetic dipolar interaction favors macroscopically degenerate ``2-in, 2-out'' configurations without any LRO, forming a magnetic analog of the water ice~\cite{bramwell:01}. Then, the classical spins are quenched into one of the degenerate ground states~\cite{castelnovo:10}. Usually, a thermal heating is required for melting the quenched spin ice. Here, we pursue an alternative possibility that quantum fluctuations melt the spin ice: the quantum entanglement among degenerate states lift the macroscopic degeneracy, suppress the spin-ice freezing, and lead to a distinct ground state. 

A realistic approach to the quantum melting of the spin ice is to choose a rare-earth ion with fewer $f$ electrons and a smaller magnetic moment, e.g., Pr$^{3+}$. In rare-eath ions with fewer $f$ electrons, the $4f$ wavefunction is spatially extended~\cite{rossat-mignod:83} and can then be largely overlapped with the O $2p$ orbitals at the O1 site (Fig.~\ref{fig:crystal} (a)) in the pyrochlore lattice. Besides, for Pr$^{3+}$ ions, the magnetic dipolar interaction, which is proportional to the square of the moment size, is reduced by an order of magnitude to 0.1~K between the nearest-neighbor sites, in comparison to that for Dy$^{3+}$ ions. Then, the superexchange interaction due to virtual $f$-$p$ electron transfers, which provides a source of the quantum nature, is expected to play crucial roles in Pr$_2TM_2$O$_7$ ($TM$: a transition metal).

Recent experiments on Pr$_2$Sn$_2$O$_7$~\cite{matsuhira:02}, Pr$_2$Zr$_2$O$_7$~\cite{matsuhira:09}, and Pr$_2$Ir$_2$O$_7$~\cite{nakatsuji:06} have shown that the Pr$^{3+}$ ion provides the $\langle111\rangle$ Ising moment described by a non-Kramers magnetic doublet. As in the spin ice, any magnetic dipole LRO is absent~\cite{matsuhira:02,matsuhira:09,nakatsuji:06,machida:09,zhou:08,maclaughlin:08}. Pr$_2$Ir$_2$O$_7$ shows a metamagnetic transition only when the magnetic field is applied in the [111] direction~\cite{machida:09}, indicating the ice-rule formation due to a ferromagnetic coupling $J\sim1.4$ K~\cite{machida:09}. 
On the other hand, the Curie-Weiss temperature $T_{CW}$ is antiferromagnetic for the zirconate~\cite{matsuhira:09} and iridate~\cite{nakatsuji:06}, unlike the spin ice. The stannate shows a significant level of low-energy short-range spin dynamics~\cite{zhou:08}, which is absent in the classical spin ice. Furthermore, the iridate shows the Hall effect at zero magnetic field without magnetic dipole order~\cite{machida:09}, suggesting an onset of a chiral spin liquid~\cite{wen:89} at a temperature $\sim J$ due to quantum fluctuations.

In this Letter, we derive the realistic effective model for Pr $4f$ moments on the pyrochlore lattice. It contains appreciable quantum nature leading to a cooperative ferroquadrupolar ground state, accompanied by crystal symmetry lowering from cubic to tetragonal and a frustration in the chirality ordering. 
Our scenario explains unusual magnetic properties observed in Pr$_2TM_2$O$_7$ suggesting a dynamically fluctuating spin ice~\cite{zhou:08,machida:09,maclaughlin:08}.

\begin{figure}[t]
\begin{center}
\includegraphics[width=\columnwidth]{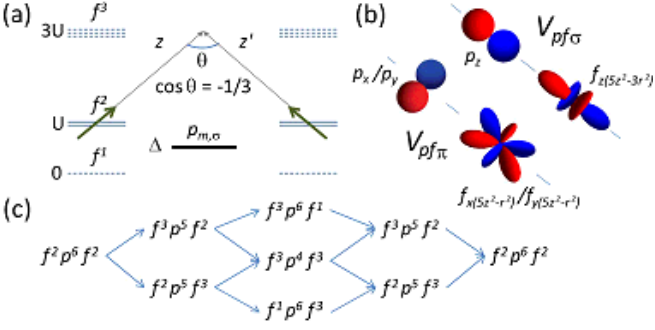}
\end{center}
\caption{(Color online) (a) Local level scheme for $f$ and $p$ electrons, and the local quantization axes $\vec{z}_{\bm{r}}$ and $\vec{z}_{\bm{r}'}$. (b) $f$-$p$ transfer integrals.
%; $V_{pf\sigma}$ between $p_z$ and $f_{(5z^2-3r^2)z}$ orbitals, and $V_{pf\pi}$ between $p_x$/$p_y$ and $f_{x(5z^2-r^2)}$/$f_{y(5z^2-r^2)}$. 
(c) Virtual hopping processes. $n$ ($n'$) and $\ell$ in the state $f^np^\ell f^{n'}$ represent the number of $f$ electrons at the Pr site $\bm{r}$ ($\bm{r}'$) and that of $p$ electrons at the O1 site.}
\label{fig:perturbation}
\end{figure}

We start with $f^2$ configurations for Pr$^{3+}$ forming the tetrahedron centered at the O$^{2-}$ ion (O1) in Pr$_2TM_2$O$_7$ (Fig.~\ref{fig:crystal}). The $LS$ coupling gives the ground-state manifold ${}^3H_4$. Each Pr$^{3+}$ ion is placed in a crystalline electric field (CEF) which has the $D_{3d}$ symmetry about the $\langle111\rangle$ direction toward the O1 site. It is useful to define the local quantization axis $\vec{z}_{\bm{r}}$ as this direction, as well as $x$ and $y$ axes as $\vec{x}_{\bm{r}}$ and $\vec{y}_{\bm{r}}$ depicted in Figs.~\ref{fig:crystal} (a) and (b). The CEF favors $J^z=\pm 4$ configurations for the total angular momentum, which are linearly coupled to $J^z=\pm1$ and $\mp2$ because of the $D_{3d}$ CEF~\cite{machida:phd}. This leads to the atomic non-Kramers magnetic ground doublet 
\begin{equation}
  |\sigma^z\rangle=\alpha|J^z=4\sigma^z\rangle+\beta\sigma^z|J^z=\sigma^z\rangle-\gamma|J^z=-2\sigma^z\rangle,
\label{eq:local}
\end{equation}
with small real coefficients $\beta$ and $\gamma$ as well as $\alpha=\sqrt{1-\beta^2-\gamma^2}$. The pseudospin $\sigma^z=\pm$ represents the direction of the Ising (``in'' or ``out'') magnetic dipole moment, in contrast to the case of a {\it nonmagnetic} doublet labeled by the atomic quadrupole moment~\cite{cox:87} in materials having other CEF symmetries, PrFe$_4$P$_{12}$~\cite{aoki:02}, UPt$_3$~\cite{upt3}, and UPd$_2$Al$_3$~\cite{upd2al3}. For Pr$_2$Ir$_2$O$_7$, the first excited crystal-field level is a singlet located at 168~K and the second is a doublet at 648~K~\cite{machida:phd}. They are similarly large for Pr$_2$Sn$_2$O$_7$~\cite{zhou:08}. These energy scales are two orders of magnitude larger than our relevant energy scale $J\sim1.4$~K. Hence we neglect these CEF excitations.

\begin{figure}[t]
\begin{center}
\includegraphics[width=7.5cm]{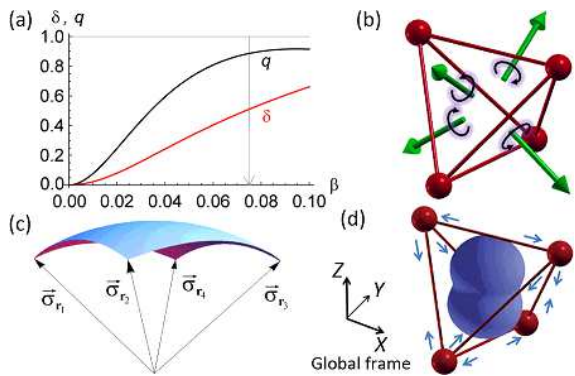}
\end{center}
\caption{(Color online) (a) Coupling constants $\delta$ and $q$ versus $\beta(=\gamma/3)$. The arrow points to the experimentally estimated value of $\beta$~\cite{machida:phd}. (b) Outward normal vectors (green arrows) of the surfaces of the tetrahedron, used to define the chirality $\kappa_T$. (c) Solid angle subtended by four pseudospins $\vec{\sigma}_{\bm{r}_i}$. (d) Distribution of the tetrahedral magnetic moment $\vec{M}_T$ in the cooperative ferroquadrupolar state ($\langle Q_T^{ZZ}\rangle>0$). The arrows represent the lattice deformation linearly coupled to $Q_T^{ZZ}$. 
%The global axes $X$, $Y$, and $Z$ are also shown.
}
\label{fig:coupling}
\end{figure}

Now we derive the effective Hamiltonian through the fourth-order strong-coupling perturbation theory. Virtual local $f^1$ and $f^3$ states have an energy gain of the Coulomb repulsion $U$ and cost of $2U$, respectively (Fig.~\ref{fig:perturbation} (a)), where the $LS$ coupling has been ignored in comparison with $U$. Creating a virtual $p$ hole decreases the energy by the $p$ electron level $\Delta$ compared with the $f^1$ level. 
The $f$-$p$ electron transfer is allowed only within the orbital $l_z=0$ and $\pm1$ manifolds, whose amplitudes are given by Slater-Koster parameters $V_{pf\sigma}$ and $V_{pf\pi}$~\cite{sharma:79}, respectively (Fig.~\ref{fig:perturbation} (b)).
The local coordinate frames for the nearest-neighbor Pr sites are crucially different; for instance, $\vec{z}_{\bm{r}}\cdot\vec{z}_{\bm{r}'}=-1/3$. The perturbation expansion in $V_{pf\sigma}$ and $V_{pf\pi}$ is then carried out by taking into account the different local coordinate frames and the virtual processes (Fig.~\ref{fig:perturbation} (c)).
The projection of this superexchange Hamiltonian onto the subspace of doublets (Eq.~(\ref{eq:local})) leads to the pseudospin-$1/2$ Hamiltonian;
\begin{eqnarray}
  {\cal H}_{\mathrm{eff}}&=&J\sum_{\langle\bm{r},\bm{r}'\rangle}^{\mathrm{n.n}}\left[\sigma_{\bm{r}}^z\sigma_{\bm{r}'}^z+2\delta\left(\sigma_{\bm{r}}^+\sigma_{\bm{r}'}^-+\sigma_{\bm{r}}^-\sigma_{\bm{r}'}^+\right)
    \right.
    \nonumber\\
    &&\left.+2q\left(e^{i\varphi_{\bm{r},\bm{r}'}}\sigma_{\bm{r}}^+\sigma_{\bm{r}'}^++h.c.\right)\right],
  \label{eq:H_eff}
\end{eqnarray}
with $\sigma^\pm_{\bm{r}}\equiv(\sigma^x_{\bm{r}}\pm i\sigma^y_{\bm{r}})/2$ and $(\sigma^x_{\bm{r}},\sigma^y_{\bm{r}},\sigma^z_{\bm{r}})=\vec{\sigma}_{\bm{r}}\cdot(\vec{x}_{\bm{r}},\vec{y}_{\bm{r}},\vec{z}_{\bm{r}})$,
where $\vec{\sigma}_{\bm{r}}$ represents the Pauli matrix for the pseudospin at the site $\bm{r}$. We have adopted the simplest gauge where the phase $\varphi_{\bm{r},\bm{r}'}$ takes $0$, $2\pi/3$, $-2\pi/3$ depending on the color of the bond directions shown in Fig.~\ref{fig:crystal} (c), by rotating the $x$ and $y$ axes by $\pi/12$ from those shown in Figs.~\ref{fig:crystal} (a) and (b).  This phase can not be fully gauged away, because of the noncollinearity of the $\langle111\rangle$ magnetic moments and the three-fold rotational invariance of $(\bm{r}, \vec{\sigma}_{\bm{r}})$ about the [111] axes. Only $\sigma^z_{\bm{r}}$ contributes to the {\em magnetic dipole moment}  $J^z_{\bm{r}}$, while $\sigma^{x,y}_{\bm{r}}$ the {\em atomic quadrupole moment} $J^z_{\bm{r}}J^{x,y}_{\bm{r}}$, as can be shown by direct calculations. For a realistic case $-0.37\lesssim V_{pf\pi}/V_{pf\sigma}\lesssim-0.02$, the Ising coupling $J$ between the nearest-neighbor pseudospins is found to be positive, i.e., antiferroic. This indicates the ``ferromagnetic'' coupling between the nearest-neighbor $4f$ magnetic moments because of the tilting of the local $z$ axes, $\vec{z}_{\bm{r}}\cdot\vec{z}_{\bm{r}'}=-1/3$. Then, it can provide a source of the ice-rule formation. The $D_{3d}$ CEF creates two additional quantum-mechanical interactions; the pseudospin-exchange and pseudospin-nonconserving terms. Their coupling constants $\delta$ and $q$ are insensitive to $U/V_{pf\sigma}$ and $\Delta/V_{pf\sigma}$ but strongly depends on $\beta$ and $\gamma$. Figure~\ref{fig:coupling} (a) shows $\delta$ and $q$ as functions of $\beta$ for the trigonal CEF with by keeping the ratio $\gamma/\beta=3$. Henceforth, we adopt rough estimates $U/V_{pf\sigma}=5$, $\Delta/V_{pf\sigma}=4$, and $V_{pf\pi}/V_{pf\sigma}=-0.3$ from first-principles calculations (published elsewhere), and $\beta=7.5\%$ and $\gamma=3\beta$ from the CEF analysis based on inelastic neutron-scattering experiments~\cite{machida:phd}. Then, we obtain $\delta\sim0.51$ and $q\sim0.89$, indicating the appreciable quantum nature. The two couplings play crucial roles in inducing a cooperative ferroquadrupolar order instead of the classical spin ice~\cite{bramwell:01} or the U(1) spin liquid~\cite{hermele:04}.

A mean-field analysis~\cite{reimers:91} on Eq.~(\ref{eq:H_eff}) yields two distinct states. (i) Magnetic dipolar states characterized by a nonzero $\langle\sigma^z_{\bm{r}}\rangle$ have the lowest energy $-2J$ per tetrahedron at the wavevector $\bm{q}=\frac{2\pi}{a}(hhl)$ with $a$ being the side length of the unit cube (Fig.~\ref{fig:crystal} (c)). (ii) A quadrupolar state with a nonzero $\langle\sigma^{x,y}_{\bm{r}}\rangle$ has the energy $-2(\delta+2q)J$ at $\bm{q}=0$ for $\delta,q>0$. Thus, for $\delta+2q>1$ as in our case, the atomic quadrupole moment $\sigma^{x,y}_{\bm{r}}$ should form the LRO {\it without any dipole LRO}. However, we will show below that the ground state is further replaced with a cooperative ferroquadrupolar state because of the quantum interplay between atomic dipoles $\sigma^z_{\bm{r}}$ and quadrupoles $\sigma^{x,y}_{\bm{r}}$.

First let us solve Eq.~(\ref{eq:H_eff}) on a single tetrahedron. The similar analysis on a distinct model for Tb$_2$Ti$_2$O$_7$~\cite{gardner:99} has been employed to discuss the RVB-singlet quantum spin ice~\cite{molavian:07}. With increasing $\beta$ and thus $\gamma$ from 0, three classical levels corresponding to the ``2-in, 2-out'', ``3-in, 1-out''/``1-in, 3-out'' (${\mit\Delta}E=2J$), and ``4-in''/``4-out'' (${\mit\Delta}E=8J$) configurations are split to three doublets, three triplets, and one singlet. In our case, the ground-state manifold has the $E_g$ symmetry with the double degeneracy $\chi=\pm$ as described as
  $|\Psi_\chi^s\rangle=\frac{c_2}{\sqrt{6}}\sum_{\tau=\pm}(
e^{i\frac{2\pi}{3}\chi}|\tau X\rangle
+e^{-i\frac{2\pi}{3}\chi}|\tau Y\rangle
+|\tau Z\rangle)+c_4|4\chi\rangle$
with real coefficients $c_2$ and $c_4$. Here, the orthonormal state $|+4\rangle$/$|-4\rangle$ represents the ``4-in''/''4-out'' configuration, while $|\pm X\rangle$, $|\pm Y\rangle$, and $|\pm Z\rangle$ denote the ``2-in, 2-out'' having the net magnetic dipole moment
  $\vec{M}_T=M_0\sum{}_{\bm{r}}^T\sigma_{\bm{r}}^z\vec{z} _{\bm{r}}$,
 pointing to the $\pm X$, $\pm Y$, and $\pm Z$ directions of the global coordinate frame, respectively. We have introduced the moment amplitude $M_0=g_J\mu_B(4\alpha^2+\beta^2-2\gamma^2)\approx2.9\mu_B$ with the Land\'{e} factor $g_J=4/5$.
The sign $\chi=\pm$ represents the net pseudospin chirality of the tetrahedron, 
  $\kappa_T=\frac{1}{2}\sum{}_{\bm{r}_1,\bm{r}_2,\bm{r}_3}^T\vec{\sigma}_{\bm{r}_1}\cdot\vec{\sigma}_{\bm{r}_2}\times\vec{\sigma}_{\bm{r}_3}$,
through the relation $\langle\Psi_\chi^s|\kappa_T|\Psi_{\chi'}^s\rangle=\sqrt{3}c_2^2\chi\delta_{\chi,\chi'}$. Here, the summation over the sites $\bm{r}_1,\bm{r}_2,\bm{r}_3$ on the tetrahedron $T$ is taken so as they appear counterclockwise about the outward normal to the plane spanned by the three sites (Fig.~\ref{fig:coupling} (b)). This $\kappa_T$ gives the solid angle subtended by the four pseudospins (Fig.~\ref{fig:coupling} (c)). 
Note that the ``2-in, 2-out'' singlet state with the $A_{1g}$ symmetry~\cite{molavian:07}, $\sum_{\tau=\pm}(|\tau X\rangle+|\tau Y\rangle+|\tau Z\rangle)/\sqrt{6}$, is located at a high energy ${\mit\Delta}E\sim7J$. The triply degenerate first excited states consist of only ``3-in, 1-out'' and ``1-in, 3-out'', and are located at ${\mit\Delta}E\sim J$.
In fact, any eigenstate of the single-tetrahedron Hamiltonian is described by either ``3-in, 1-out'' and ``1-in, 3-out'' configurations or ``2-in, 2-out'' and ``4-in''/``4-out'' configurations. Therefore, quantum effects of creating ``3-in, 1-out'' and ``1-in, 3-out'' configurations from the ``2-in, 2-out'' can not be taken into account in the single-tetrahedron analysis.

To overcome this drawback, we numerically solve the model for the 16-site ($N=16$) cubic cluster with the periodic boundary condition (Fig.~\ref{fig:crystal} (c)). It is found that the ground states have a six-fold degeneracy labeled by the inversion ($I$) even(+)/odd(-) and the wavevector $\bm{k}_X=(\frac{2\pi}{a},0,0)$, $\bm{k}_Y=(0,\frac{2\pi}{a},0)$, or $\bm{k}_Z=(0,0,\frac{2\pi}{a})$ with the energy $\sim -8.825J$ per tetrahedron. The states associated with $\bm{k}_i$ have a cooperative quadrupole moment defined on each tetrahedron, $\langle Q^{ii}_T\rangle=0.0387M_0^2$, where
  $Q^{ij}_T=3M^i_TM^j_T-\vec{M}^2_T\delta_{ij}$
with $i,j=X, Y, Z$.
Namely, the net magnetic moment $\vec{M}_T$ in each tetrahedron $T$ points, for instance, to the $\pm Z$ directions with a higher probability than to the $\pm X$ and $\pm Y$ (Fig.~\ref{fig:coupling} (d)). Such ferroquadrupole order spontaneously breaking the three-fold rotational invariance about the $[111]$ axes can occur in the thermodynamic limit. This ferroquadrupole moment $Q_T^{ii}$ linearly couples to a lattice vibration: the four ferromagnetic bonds and the two antiferromagnetic bonds should be shortened and expanded, respectively, leading to a crystal symmetry lowering from cubic to tetragonal accompanied by a compression in the direction of the ferroquadrupole moment (Fig.~\ref{fig:coupling} (d)). This state shows both axial alignments of magnetic dipoles and a broken translational symmetry, and can then be classified into a magnetic analog of a smectic (or crystalline) phase of liquid crystals~\cite{degennes}. Such magnetic quadrupole correlations in Pr$_2TM_2$O$_7$ could be probed by NMR experiments.

\begin{figure}
\begin{center}
\includegraphics[width=\columnwidth]{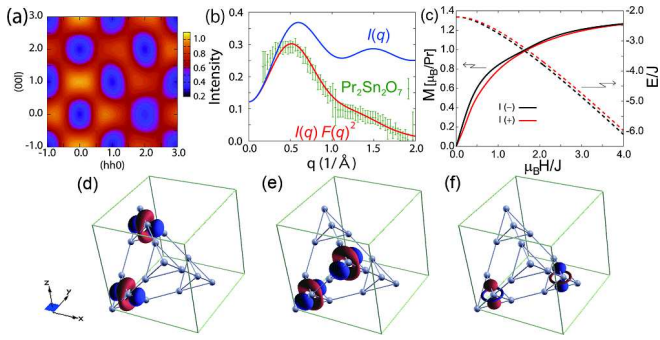}
\end{center}
\caption{(Color) (a) $S(\bm{q})/M_0^2$ constructed from the local, nearest-neighbor, and second-neighbor correlations. (b) The theoretical curve (red) $I(q)F(q)^2$ with the form factor $F(q)$ and the experimental data (green) of the powder neutron-scattering intensity on Pr$_2$Sn$_2$O$_7$ at 1.4~K~\cite{zhou:08}. $I(q)$ (blue curve) is the angle average of $S(\bm{q})/M_0^2$. (c) The magnetizations (left) and energies (right) per site for the $I$-odd(-) ground state and the $I$-even(+) state  under the magnetic field $\vec{H}\parallel\langle111\rangle$.
(d-f) Quadrupole correlations $\langle Q^{ii}_TQ^{jj}_{T'}\rangle$ between the tetrahedrons $T$ and $T'$ displaced by $\bm{r}=\bm{r}_X$ (d), $\bm{r}_Y$ (e), and $\bm{r}_Z$ (f) in the cooperative ferroquadrupolar state with the wavevector $\bm{k}_Z$ and $\langle Q_T^{ZZ}\rangle\ne0$. The matrix $\langle Q^{ii}_TQ^{jj}_{T'}\rangle$ in $i,j$ has been diagonalized to yield two orthogonal forms of quadrupoles, ${\cal Q}_{\bm{r}\mu}=\sum_i\lambda_{\bm{r}\mu}^iQ_T^{ii}$ ($\mu$=1,2). The shape of ${\cal Q}_{\bm{r}1}$ showing the dominant correlation amplitude is shown. Red/blue regions represent positive/negative values of ${\cal Q}_{\bm{r}1}$.}
\label{fig:MH-q}
\end{figure}

Next, we calculate the magnetic dipole correlation,
$S(\bm{q})=\frac{M_0^2}{N}\sum_{\bm{r},\bm{r}',i,j}(\delta_{ij}-\frac{q_iq_j}{|\bm{q}|^2})z_{\bm{r}}^i z_{\bm{r}'}^j\langle \sigma^z_{\bm{r}}\sigma^z_{\bm{r}'}\rangle_{\mathrm{ave}}e^{i\bm{q}\cdot(\bm{r}-\bm{r}')}$,
averaged over the six-fold degenerate ground states. This quantity is relevant to the neutron-scattering intensity integrated over the low-energy region below the crystal-field excitations from the atomic ground doublet Eq.~(\ref{eq:local}). Note that neutron spins do not couple to $\sigma_{\bm{r}}^\pm$ which represents the atomic quadrupole. Figure~\ref{fig:MH-q} (a) shows the profile of $S(\bm{q})$ for $\bm{q}=\frac{2\pi}{a}(hhl)$. It exhibits intense peaks at $(001)$ and $(003)$, weaker peaks at $(\frac{3}{4}\frac{3}{4}0)$, and the minimum at $(000)$, as in the dipolar spin ice~\cite{bramwell:01}, though the peaks are broadened by the quantum fluctuations in this ferroquadrupolar state. Besides, the nonzero $q$ term in the Hamiltonian partially violates the ice rule and eliminates the pinch-point singularity~\cite{isakov:04,henley:05} observed at (111) and (002) in the spin ice~\cite{fennell:07}, which should be examined by large system-size calculations.
Note that our magnetic profile reproduces powder neutron-scattering results on Pr$_2$Sn$_2$O$_7$~\cite{zhou:08} that reveal the enhanced low-energy short-ranged intensity at $q\sim\frac{2\pi}{a}\sim0.5$~\AA$^{-1}$ with a shoulder at $q\sim\frac{6\pi}{a}$ (Fig.~\ref{fig:MH-q}(b)).
 This experiment also shows a saturation of the quasielastic peak width $\sim0.1$~meV~$\sim J$ at 0.2~K~\cite{zhou:08}. Such large spin relaxation rate can be attributed to the appreciable quantum nature; large $\delta$ and $q$ in Eq.~(2). 
These agreements support our scenario of the quantum melting of a spin ice.

Now we concentrate on the ground states having the quadrupole moment $\langle Q_T^{ZZ} \rangle>0$ and the associated wavevector $\bm{k}_Z$. The magnetic susceptibility is finite in the ferroquadrupolar state, as seen from the slope of the magnetization curve along the $\langle111\rangle$ direction in Fig.~\ref{fig:MH-q} (c). This indicates a negative $T_{CW}$ as found in Pr$_2$Zr$_2$O$_7$~\cite{matsuhira:09} and Pr$_2$Ir$_2$O$_7$~\cite{machida:09}, and the absence of an internal magnetic field as in Pr$_2$Ir$_2$O$_7$~\cite{maclaughlin:08}. The magnetic field lifts the degeneracy due to the $I$ symmetry. The ground-state ($I$-odd) magnetization shows a small step or dip around $\mu_BH/J\sim1.5$, in comparison with that of the $I$-even excited state. This indicates that the structure develops upon cooling. These agree with the experimental observation on Pr$_2$Ir$_2$O$_7$; $M\sim 0.8\mu_B$ at the metamagnetic transition $\mu_BH_c/J\sim1.3$ with $J\sim1.4$~K~\cite{machida:09}.

Finally we spatially resolve the multipolar correlations within the cubic unit cell. Figures~\ref{fig:MH-q} (d), (e), and (f) represent quadrupole correlations $\langle Q^{ii}_TQ^{jj}_{T'}\rangle$ between the tetrahedrons $T$ and $T'$ displaced by $\bm{r}_X=(0,\frac{a}{2},\frac{a}{2})$, $\bm{r}_Y=(\frac{a}{2},0,\frac{a}{2})$, and $\bm{r}_Z=(\frac{a}{2},\frac{a}{2},0)$, respectively. There exist dominant ferroquadrupolar correlations shown in Figs.~\ref{fig:MH-q} (d) and (e), both of which favor ferroquadrupole moments along the $Z$ direction. They prevail over subdominant antiferroquadrupole correlations shown in Fig.~\ref{fig:MH-q} (f), and are responsible for the ferroquadrupole order $\langle Q_T^{ZZ}\rangle\ne0$. On the other hand, the chirality correlation $\langle\kappa_T\kappa_{T'}\rangle$ is weakly ferrochiral between the tetrahedrons shown in Figs.~\ref{fig:MH-q} (d) and (e), while it is strongly antiferrochiral between those shown in Fig.~\ref{fig:MH-q} (f). This points to a geometrical frustration suppressing the chirality LRO in each fcc sublattice of the diamond lattice formed by the tetrahedrons. Further studies are required for examining a possibility of a chiral spin liquid~\cite{wen:89}. The broken time-reversal symmetry without magnetic dipole LRO, reported in Pr$_2$Ir$_2$O$_7$~\cite{machida:09}, might be detected even in  insulating magnets such as Pr$_2$Zr$_2$O$_7$ and Pr$_2$Sn$_2$O$_7$ through magneto-optical Kerr-effect measurements.

The proposed scenario of the quantum melting of the spin ice explains magnetic properties observed in Pr$_2TM_2$O$_7$. Effects of coupling of localized $f$-electrons to conduction electrons on the transport properties are left for a future study. The orbital motion of conduction electrons can flip the pseudospin-$1/2$. This could be an origin of the resistivity minimum observed in Pr$_2$Ir$_2$O$_7$~\cite{nakatsuji:06}. 

The authors thank S. Nakatsuji, Y. Machida, Y. B. Kim, K. Matsuhira, and D. MacLaughlin for discussions. The work was supported by Grants-in-Aid for Scientific Research under No. 19052006, 20029006, and 20046016 from the MEXT of Japan and 21740275 from the JSPS.

\end{document}